# Watt-level single-cavity dual-comb laser at 2.1 µm wavelength

**MYKYTA REDKIN,[1] YICHENG WANG,[1*] MARCEL VAN DELDEN,[2] ROBIN LÖSCHER,[1] OLIVER NAGY,[1] SERGEI TOMILOV,[1] AND CLARA J. SARACENO[1,3]**

[1]*Photonics and Ultrafast Laser Science, Ruhr-Universität Bochum, Universitätsstrasse 150, 44801 Bochum, Germany*
[2]*Electronic Circuits (EST), Ruhr University Bochum, Universitätsstrasse 150, 44801 Bochum, Germany*
[3]*Research Center Chemical Science and Sustainability, University Alliance Ruhr, 44801 Bochum, Germany*
**yicheng.wang@ruhr-uni-bochum.de*

**Compact ultrafast laser combs around the 2 µm atmospheric transmission window are of growing interest for direct spectroscopy and as drivers for nonlinear conversion to otherwise hard-to-reach spectral regions. Dual-comb systems in this region remain very scarce: existing implementations rely mostly on nonlinear conversion, resulting in complex systems with poor conversion efficiency and overall low average and peak power. In this paper, we report on a watt-level dual-comb laser directly emitting at 2.1 µm. The system is based on a single-cavity, soft-aperture Kerr-lens mode-locked (KLM) laser using a broadband, holmium (Ho)-doped $CaAlGdO_4$ (CALGO) gain crystal. The laser delivers a total output power exceeding 2 W (1 W per comb) with a sub-100-fs pulse duration (50-nm bandwidth) at a repetition rate of 96.5 MHz, corresponding to a pulse energy of 10 nJ and a peak power of 100 kW per comb. To assess the source's potential for future spectroscopic applications, we characterize the free-running dual-comb in detail for frequency stability, phase noise, and relative intensity noise (RIN), and benchmark achievable spectral resolution by measuring spectral modulations imposed by Fabry–Pérot etalons. Our system combines the compactness, high efficiency, and passive mutual coherence of a single-cavity oscillator with watt-level average power at 2.1 µm, making it a promising system for environmental sensing in the atmospheric transmission window, or as a driver for nonlinear frequency conversion, for example, to the MIR and THz regions.** https://doi.org

## 1. INTRODUCTION

Dual-comb systems have become an increasingly mature and versatile measurement platform in recent years [1,2] with applications in molecular spectroscopy that demand high resolution and rapid acquisition [3], precise distance measurements [4,5] and optical frequency calibration [6]. In all these applications, the achievable performance is set by the properties of the comb sources, their mutual coherence, noise, and available output parameters.

This has motivated the development of dual-comb architectures based on fiber, semiconductor, microresonator and solid-state lasers [7–11]. Among these, mode-locked solid-state lasers stand out: gain media with long upper-state lifetimes support particularly low noise and give access to high average powers [12,13] and often to clean soliton-type pulses. These advantages are essential for numerous applications where, for example, high powers and clean pulses are required for efficient frequency conversion to other spectral regions as well as for measurements with significant optical path losses, such as open-path atmospheric sensing. Low noise improves the achievable resolution and allows coherent averaging, for which the amplitude SNR scales as $\sqrt{N}$, where N is the number of averaged traces.

Advances in such mode-locked laser dual-comb spectroscopy (DCS) systems in terms of power, bandwidth, and noise performance have enabled many breakthroughs, including record-high refresh rates in THz time-domain spectroscopy [14,15], ultrasensitive biosensing [16], and long-distance open-path environmental monitoring [17]. More recently, single-cavity dual-comb solid-state lasers have attracted significant attention [11,18,19]. In these systems, two pulse trains share most of the cavity elements and circulate in a laser cavity with a small offset in repetition rate. Both combs originate from the same resonator and common-mode noise is largely suppressed, therefore high mutual coherence is obtained passively. The single-cavity approach also reduces footprint and complexity compared to the standard configuration of having two separate comb sources mutually referenced by active stabilization electronics [20].

So far, most progress in mode-locked solid-state laser-based DCS systems, in both dual- and single-cavity configurations, has been achieved in the 1-µm and 1.5-µm spectral regions [11,19,21–23], driven by the maturity of Yb- and Er-doped gain media. However, extending this performance to longer wavelengths is of growing interest. In particular, the short-wave infrared (SWIR) spectral region between 1.7 µm and 2.4 µm is particularly attractive [24,25]: high-power SWIR sources enable efficient frequency conversion and supercontinuum generation deep into the mid-infrared spectral region (MIR) - for example, to achieve dual-comb OPOs [26–28] - using non-oxide nonlinear materials whose transparency extends beyond that of oxides and which suffer less from multiphoton absorption than under 1 µm pumping [29]. Furthermore, direct spectroscopy in the SWIR atmospheric transmission window around 2.1 µm is promising for several molecular species of environmental relevance, such as carbon dioxide ($CO_2$) and nitrous oxide ($N_2O$) [30].

In this spectral region, fiber lasers are the most widely used comb sources, offering low noise and stable operation [31–34] but at average powers significantly below those accessible with bulk gain media. Whereas bulk based laser architectures have advanced substantially in this spectral region in recent years [35–40], dual-comb demonstrations remain scarce in both single- and dual-cavity architectures. Most use $Cr^{2+}$-doped II-VI laser gain media: one demonstration of a semiconductor saturable absorber mirror (SESAM) mode-locked single-cavity Cr:ZnS dual-comb system operating at 2360 nm reached 200 fs and 30 nm of bandwidth, exploiting the broad emission of $Cr^{2+}$-doped media, with 250 mW per comb at 242 MHz [41]. Another implementation is based on two lasers, which reach watt-level power [42,43] at the cost of two separate resonators and the frequency-locking electronics. Both of these systems operate at ~2.45 µm central wavelength, in a spectral region where water absorption is significantly stronger, which can be severely limiting when considering the very broad bandwidths achieved. $Cr^{2+}$-doped crystals remain the most established bulk gain media in this spectral region due to their ultrabroad gain, which supports few-cycle pulses. However, large thermal optical coefficients and microsecond upper-state lifetimes make average-power scaling and passive noise suppression more challenging.  and $Ho^{3+}$-doped materials are more favorable in both regards: high thermal conductivity and long millisecond upper-state lifetimes enable average-power scaling and excellent stability and low-noise operation due to small relaxation oscillation frequencies. In particular, $Ho^{3+}$-doped materials benefit from single-mode in-band pumping at 1.9 µm, corresponding to a low quantum defect, resulting in reduced heat load and higher power efficiency. [36] They also emit directly at 2.1 µm, within the atmospheric transmission window, which suits free-space propagation applications. However, until recently, the pulses obtained from Holmium modelocked lasers were long (typically multi-ps) and unsuited for broadband mode-locking, due to narrowband and structured emission spectra. To the best of our knowledge, the only dual-comb demonstration using a rare-earth-doped medium thus far is a single-cavity, polarization-multiplexed Tm,Ho:YLF laser delivering 580 mW in total (282 mW and 298 mW per comb) at 1.715 GHz, but with 157 ps pulses and correspondingly very narrow spectral coverage [44].

Here, we demonstrate a single-cavity dual-comb laser based on the broadband rare-earth-doped material Ho:CALGO [36]. This material combines long upper-state lifetimes, broadband emission around 2.1 µm supporting sub-100-fs pulses, and excellent thermal properties [45,46]. Recent results have confirmed its potential for broadband, high-power and high-energy lasers at 2.1 µm, with mode-locked oscillators reaching record-high average powers [47], low-noise operation [48] and GHz repetition rates [49], as well as in broadband regenerative amplification [50,51]. Using this material, we present here a single-cavity dual-comb system with a total output power exceeding 2 W (1 W per comb), generating sub-100-fs pulses with a repetition rate of 96.5 MHz, corresponding to an output pulse energy of 10 nJ per comb. Compared with previously demonstrated single-cavity dual-comb lasers beyond 2 µm, our system delivers three times the average output power per comb of the Tm,Ho:YLF system [44] and twice the spectral bandwidth of the Cr:ZnS system [41]. We present a detailed characterization of the phase noise and relative intensity noise (RIN) of both combs. The single-cavity geometry significantly reduces the integrated timing jitter of the mixed signal, relative to the individual combs. In the mode-locked regime, the RIN measurements reveal a full suppression of the relaxation-oscillation peak present in continuous-wave operation, reducing the integrated RIN below 0.1%. Furthermore, we benchmark the free running performance of the dual comb by measuring Fabry Perot etalon lines of varying linewidths.

An overview of existing technologies in the SWIR spectral region is presented in Fig. 1, including systems reviewed above and the result presented here, illustrating the advantage of our system: high output power, single-cavity design and operation in the middle of the high transmission atmospheric window, facilitating applications and power scaling.

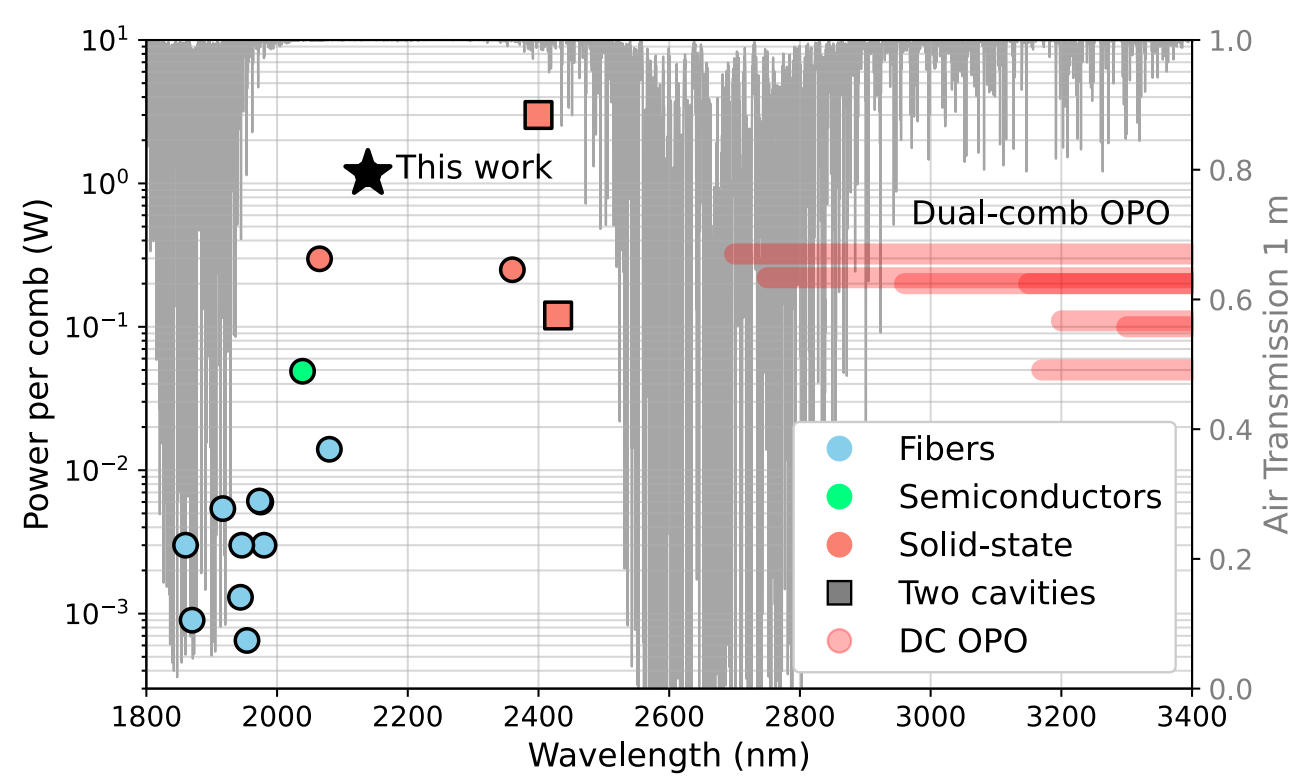

**Fig. 1.** Overview of two- [42,43] and single-cavity [24,25,31,32,41,44,52–58] dual-comb sources, together with dual-comb optical parametric oscillator schemes, operating beyond 1800 nm. DC OPO – dual-comb laser system, whose bandwidth is extended by the optical parametric oscillator [27,28,59–63].

## 2. EXPERIMENTAL SETUP

### A. Dual-comb mode-locked laser

The experimental setup of the dual-comb system is shown in Fig. 2. The laser is pumped by a commercial continuous-wave (CW) single-mode Tm-fiber laser source capable of delivering an output power up to 30 W at the wavelength of 1940 nm. The pump beam is split equally by a non-polarizing 50:50 beam splitter (BS), separated to pump two separate spots in the 3-at.% doped, a-cut 15-mm Ho:CALGO crystal. The horizontal separation between the two pumped volumes is set to 2 mm to ensure sufficient pumping with minimal mode overlap.

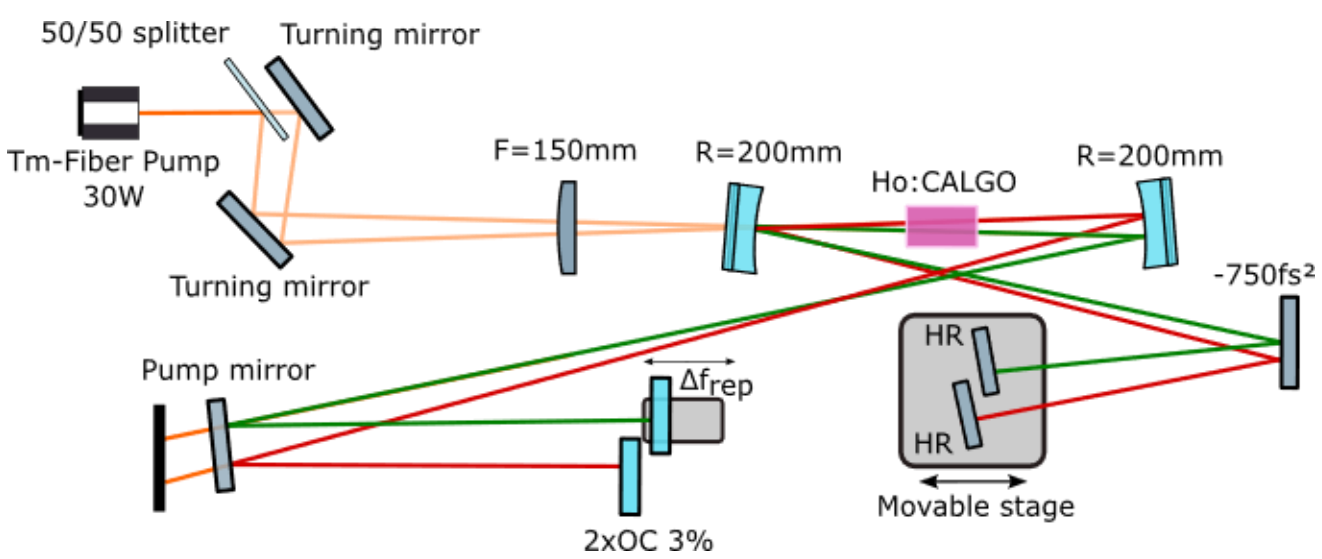

**Fig. 2.** Cavity design of the Ho:CALGO dual-comb laser system**.** HR – High Reflective mirror, OC – Output Coupler. Cavity pulse traces are indicated with corresponding colors and will be referenced further as green (comb 1) and red (comb 2).

This separation can be adjusted using a turning mirror behind the BS mounted on a translation stage. Both beams are directed onto the crystal and tightly focused to an 80-µm spot radius ($1/e^2$) using a 150-mm focal length lens. A dichroic mirror couples most of the residual pump emission out of the cavity. The laser cavity design includes two 200-mm high-reflectivity curved mirrors, one dispersive mirror introducing a round-trip group delay dispersion (GDD) of −1500 fs², two 3% output couplers (OC), and two high-reflectivity (HR) end mirrors. The Ho:CALGO crystal adds an additional round-trip GDD of −1520 fs² for σ-polarization at 2.13 µm [64], resulting in a total intracavity round-trip GDD of −3020 fs². One of the OCs is mounted on a translation stage for precise control of the difference in the repetition rates between the two combs ($\Delta f_{rep}$). On the other side of the cavity, both HR mirrors are placed on a movable stage that facilitates the simultaneous mode-locking for both laser modes [65]. The single laser design follows a similar design to Ref. 47, except for reduced pump and cavity mode radii ($1/e^2$) inside the crystal from 105 µm to 80 µm. The laser is enclosed in an air-sealed box to reduce the effect of environmental air fluctuations.

## B. Laser Characterization.

The characterization of the mode-locked laser output is shown in Fig. 3. Both laser combs are mode-locked using soft-aperture Kerr-lens mode-locking (KLM) at a total incident pump power of 17 W. The optical spectra are shown in Fig. 3(a,d), showing both combs carry a broad spectral bandwidth with full-width at half-maximum (FWHM) values of 50.2 nm and 53.3 nm, centred at 2135 nm and 2133 nm, respectively. The small offset in center wavelength originates from the quasi-three-level character of the $Ho^{3+}$ 2.1-µm transition, which shifts toward shorter wavelengths at higher inversion levels [36]. Since the two beams sample different regions of the shared crystal, slight variations in local $Ho^{3+}$ doping density and host material quality account for the measured difference.

Pulse duration measurements were performed using an intensity autocorrelator. The recorded intensity autocorrelation traces, shown in Fig. 3(b, e), demonstrate near-transform-limited soliton pulses with durations of 94 fs and 90 fs. Maximum average output powers of 1.17 W and 1.02 W were obtained for the two combs, respectively, at a laser repetition rate of 96.5 MHz, resulting in pulse energies of approximately 10 nJ per comb. The radio-frequency (RF) spectra were measured by a high-speed InGaAs photodetector with a 12-GHz bandwidth (-3 dB) and recording the signal with an RF spectrum analyzer. The RF spectrum shown in Fig. 3(f) confirms stable mode-locking of both combs across the 1-GHz measurement span, with a repetition-rate difference between combs of $\Delta f_{rep}$ = 10 kHz, as displayed in Fig. 3(c). Our setup allows us to tune this difference within a range of 0 to 772 kHz, while maintaining stable mode-locking operation and the same output parameters by changing the position of the OC of one of the combs. Fine-tuning of $\Delta f_{rep}$ can also be achieved by shifting the position of the end-mirrors translation stage, due to the slight incidence angle difference, comparable to what is commonly achieved using a biprism operation [11].

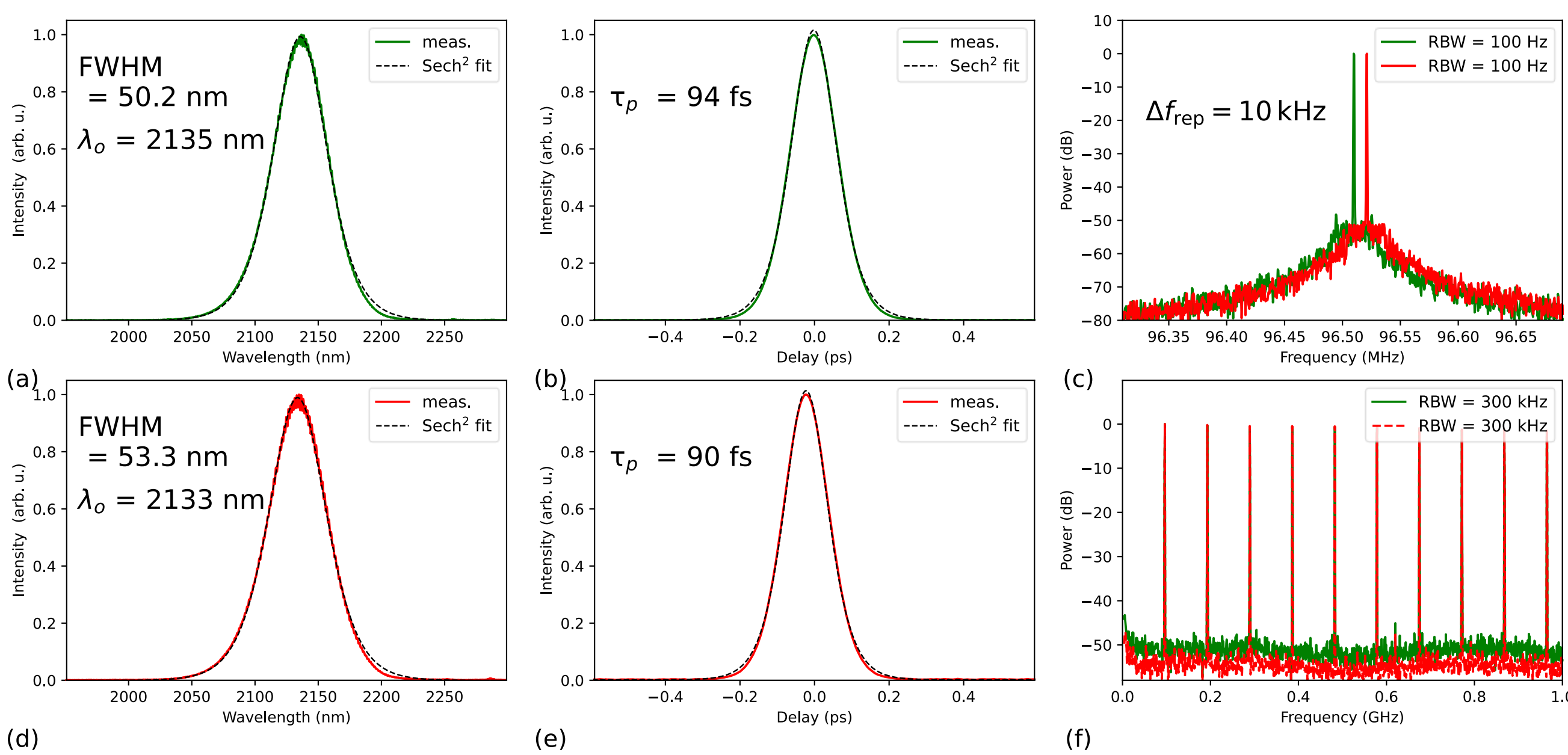

**Fig. 3.** Characterization of the dual-comb laser (a), (d) Laser spectra fitted by sech² function, showing soliton modelocking operation with more than 50 nm FWHM at a center wavelength 2135 nm and 2133 nm, respectively (b),(e) Autocorrelation of the two combs fitted with a sech² fit (c) RF spectra of the first beat note with $\Delta f_{rep}$ of 10 kHz and (f) RF spectra for two output beams indicating clean mode-locking for both combs.

In order to evaluate the long term stability of the laser, we monitored the repetition rates of both combs for a time-period of 9 hours, as shown in Fig. 4. During this test, the OC transmission was set to 1%, yielding 400 mW of average output power per comb. Lower OC transmission in this configuration results in the highest intracavity power of the system (40 W, compared with 34 W in the 1 W configuration). Increase in thermal load on the cavity optics enhances thermally induced cavity-length fluctuations and repetition-rate drift, making this a worst-case condition for long-term stability. In order to evaluate the influence of environmental factors on these long-term drifts, a sensor inside the laser cavity box measured humidity simultaneously to these $f_{rep}$ drift measurements. The measurements show an absolute drift in the laser repetition rate of about 200 Hz for both combs and a relative drift of up to 10 Hz between the two combs. After comparing the obtained data with the relative humidity inside the laser cavity, one can see an anticorrelation (note the inverted axis for humidity) between changes in the repetition rates of both combs and variations in the humidity. We attribute this behavior to the strong absorption of water vapor at the pump wavelength of 1940 nm delivered by the Tm-doped fiber laser. Atmospheric water vapor introduces significant absorption bands in the 1.9–2.0 µm spectral region [66], and variations in humidity directly affect the pump power delivered to the Ho:CALGO crystal.

Although an absolute drift is present, both combs show a similar response to environmental changes. The $\Delta f_{rep}$ value averaged during this measurement is around 156 Hz, while its standard deviation is only 0.2 Hz on a minute timescale, which is the relevant timescale for high-speed data acquisition. Over the full measurement, this value increases to about 2 Hz, but the increase is caused by the slow environmental drift of roughly 0.7 Hz/h. This stability is comparable to other free-running dual-comb systems [67–69] used for spectroscopy applications. At the time of the experiment, the primary goal was to test the system's raw operational capabilities without additional atmospheric isolation or cavity purging. This study confirms that humidity control is a meaningful consideration for achieving long-term stability, however, it shows no need for active frequency stabilization due to common-mode correlations.

Overall, the characterization results of this system highlight the long-term robustness of the mode-locking mechanism and the broad gain profile of the Ho:CALGO crystal, which supports the generation of sub-100-fs pulses with watt-level output power at the high repetition rate of sub-100-MHz.

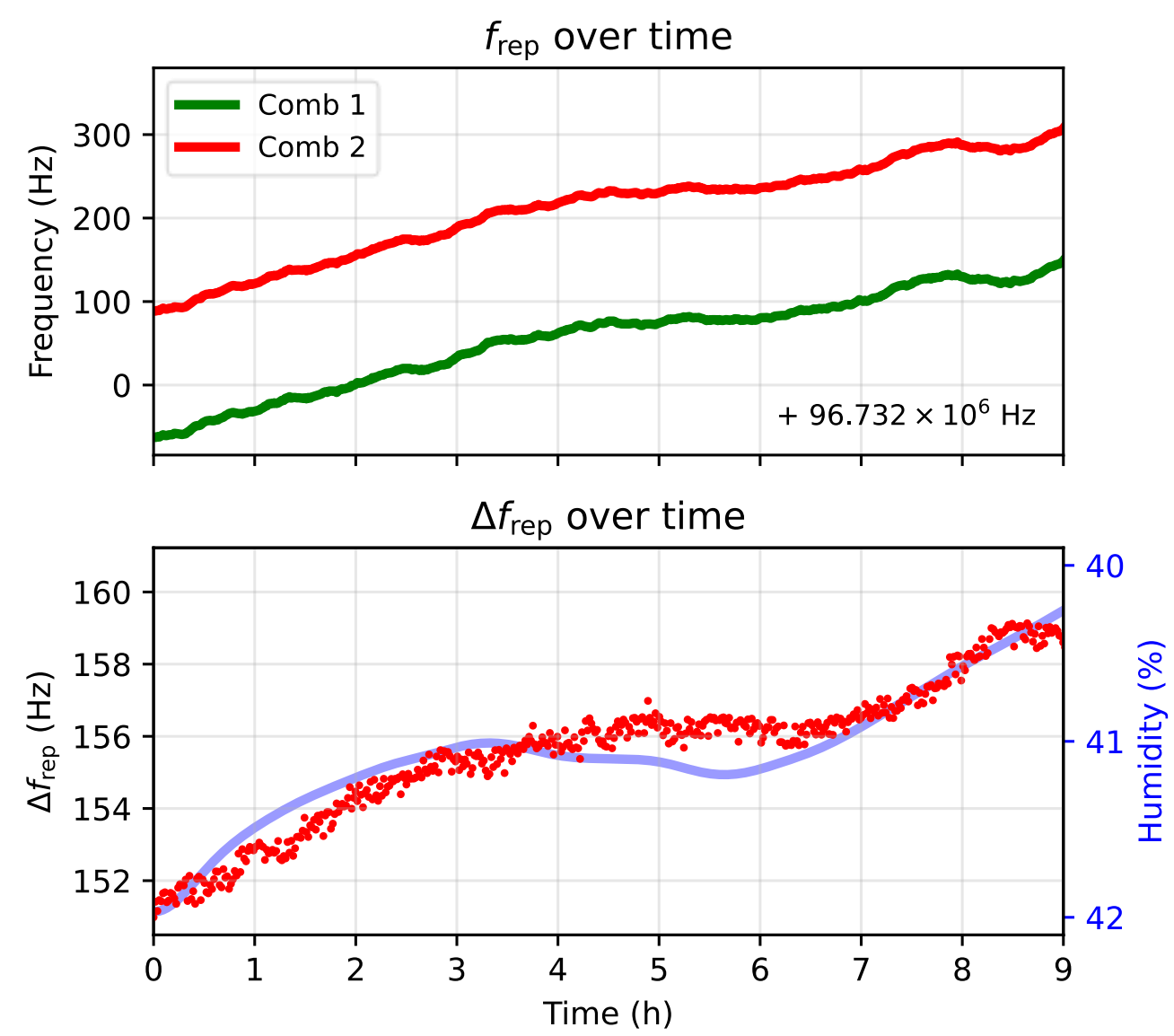

**Fig. 4.** Long-term measurement of the repetition rate drifts of two laser outputs according to the environment humidity changes.

## 3. LASER NOISE CHARACTERIZATION

One of the key advantages of the single-cavity dual-comb system is the potential coherence of the noise between the two combs. This arises from the two combs sharing the same generation medium, the same mechanical fluctuations acting on optical cavity elements, and the same intracavity environment, leading to common-mode noise suppression. This results in lower phase noise and, consequently, lower timing jitter of the dual-comb beat signal [21,70]. In this section, we characterize the phase noise and RIN properties of both comb outputs.

### A. Phase noise

For the noise characterization, the 3% OCs were reinstalled, restoring the nominal operating point with output powers of 1.05 W (Comb 1) and 0.95 W (Comb 2), which were used throughout the rest of this work. The slight deviation from the output powers reported in Section 2.B results from the cavity realignment, which shifted the mode to a different region of the inhomogeneously doped crystal. Two approaches were used to measure the phase noise and are shown schematically in Fig 5: (a) individual pulse train detection, where each comb is measured independently to obtain its absolute phase noise spectrum, and (b) electrical mixing, where the two comb signals with a set repetition rate difference ($\Delta f_{rep}$) are detected on separate fast photodiodes and subsequently mixed electrically to isolate the uncorrelated noise components.

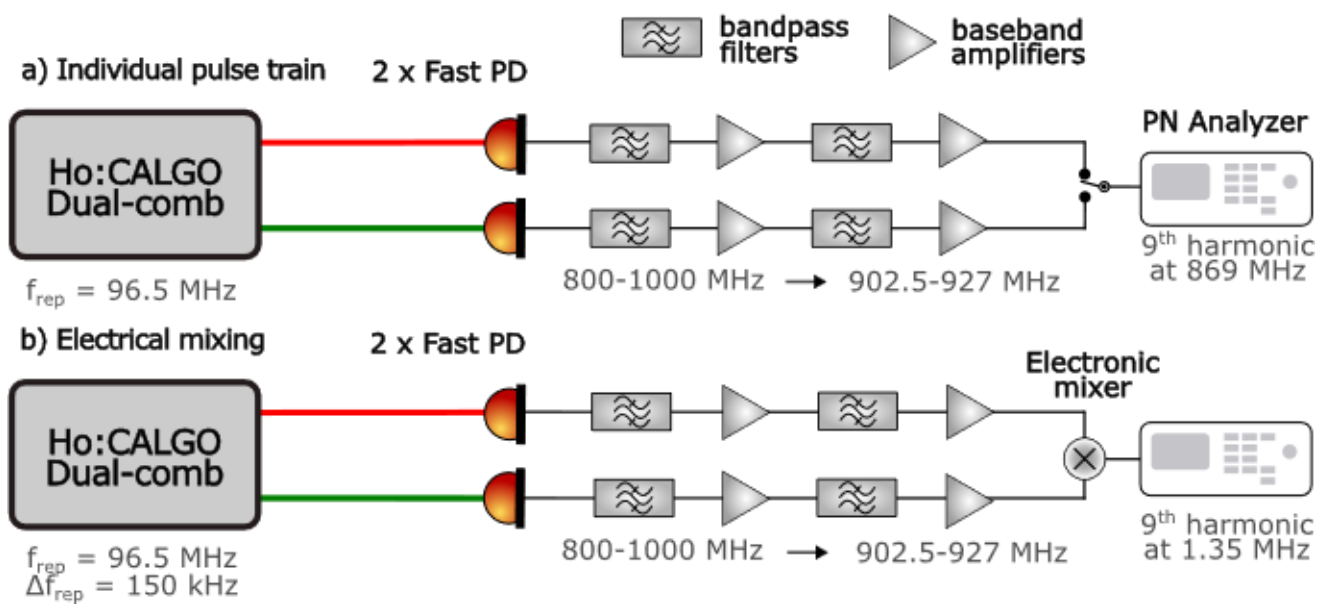

**Fig.5.** Schematic of the setup for phase noise characterization. The measurement of (a) individual combs and (b) electrically mixed beating signal. PN analyzer – phase noise analyzer.

To improve sensitivity, we performed phase noise measurements on a higher harmonic of the repetition rate rather than the fundamental. The harmonic number was chosen to maximize the signal-to-noise ratio (SNR) achievable with the available bandpass filters (BPFs) and baseband amplifiers.

The detected RF signal from each comb was first filtered by a pair of 800–1000 MHz BPFs with excellent stopband rejection to isolate the spectral region of interest and then amplified using broadband (0.05 - 10 GHz) amplifiers. A second pair of narrower BPFs (902.5 - 927 MHz) was used to suppress the remaining harmonics and spurious beat notes. As the band of these filters lies between the 9$^{th}$ and 10$^{th}$ harmonics of $f_{rep}$, the 9$^{th}$ harmonic at ~869 MHz was retained, providing the higher SNR after filtering. All phase noise data reported below were therefore acquired at the 9$^{th}$ harmonic and referred to the fundamental carrier by subtracting $20{\cdot}\log_{10}(9) = 19.1$ dB.

The $\Delta f_{rep}$ was set to 150 kHz so that the beat at the 9th harmonic, $9{\cdot}\Delta f_{rep} = 1.35$ MHz, exceeded the 1 MHz minimum carrier frequency of the phase noise analyzer (Rohde & Schwarz, FSWP50). Increasing $\Delta f_{rep}$ further requires a larger longitudinal offset between the two output couplers, which slightly modifies the cavity geometry. Both arms nevertheless share the same mechanical conditions, and both comb paths between the crystal and the output couplers are collimated, so the Kerr-lens mode-locking conditions are preserved across the tuning range. We therefore expect the noise properties to remain comparable at higher values of $\Delta f_{rep}$.

The phase noise of the two individual combs and of the mixed signal is shown in Fig. 6, together with the corresponding integrated timing jitter (right axis). In the region of 1 to 10 Hz, the two traces separate by up to 5 dB. We attribute this residual asymmetry to non-common-path contributions outside the cavity, which can be eliminated by mechanically isolating the detecting photodiodes with identical vibration-damping polymer mounts. At higher offset frequencies, between 10 Hz and 1 kHz, the two combs closely reproduce the same series of discrete spurs, confirming that mechanical perturbations act on both intracavity elements similarly.

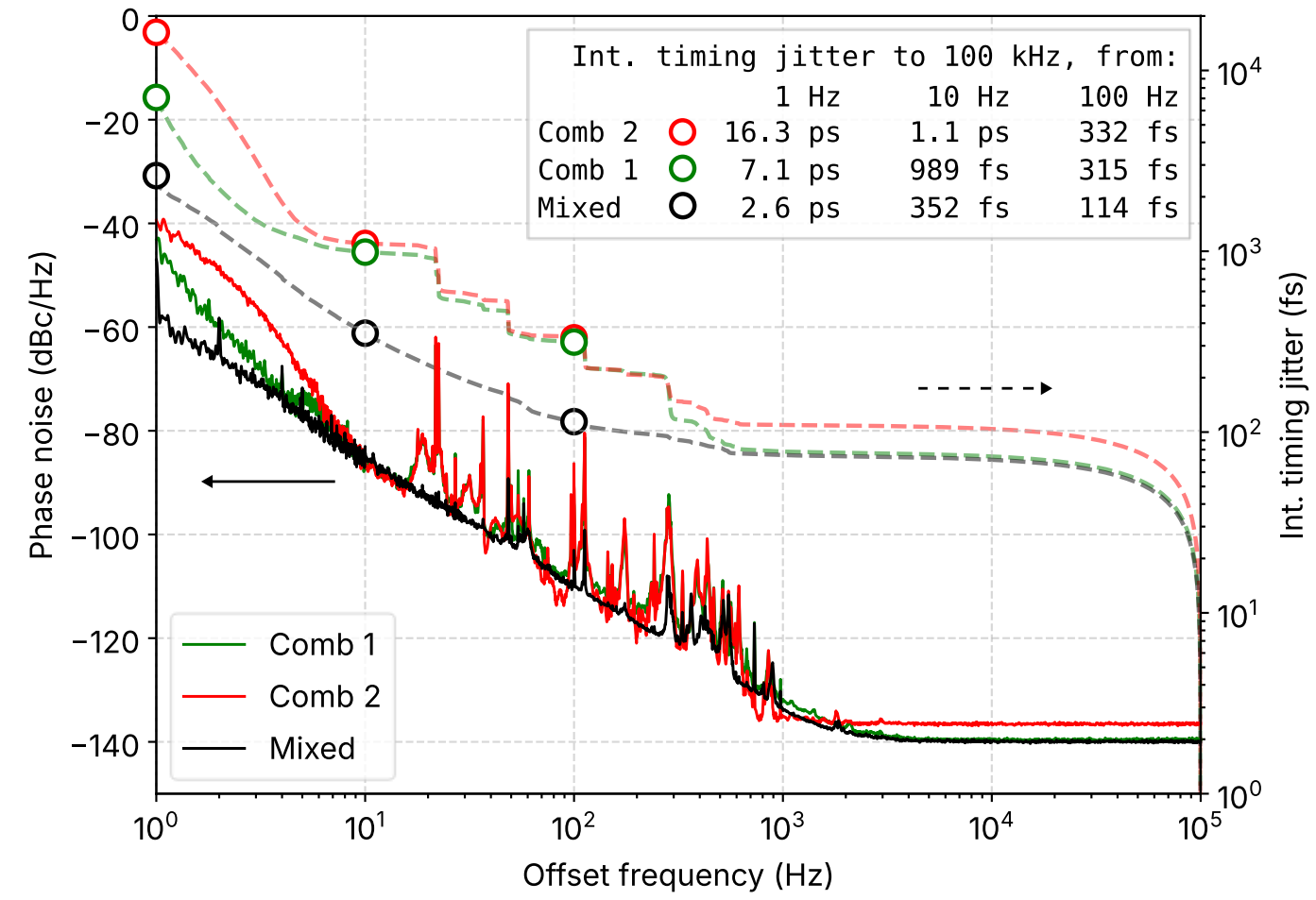

**Fig. 6.** Phase noise characteristics of individual comb at the 9$^{th}$ harmonic of the carrier. Uncorrelated phase noise (mixed) is the result of mixing two pulse trains electrically with a subsequent common mode suppression. The integrated timing jitter is shown in dashed lines, keeping the corresponding colour code.

Above 1 kHz both spectra flatten at −140 dBc/Hz and −136 dBc/Hz. This is the noise floor of the measurement setup, that limits the detection. The mixed signal lies below the individual traces, highlighting the common-mode rejection provided by the single-cavity geometry. Most of the discrete spurs present in each individual spectrum are absent in the mixed trace, confirming successful suppression of coherent perturbations. Above 1 kHz, all traces approach the measurement noise floor, so the common-mode suppression can no longer be resolved there. Integrated from 1 Hz to 100 kHz, Comb 1 and Comb 2 exhibit a timing jitter of 7.1 ps and 16.3 ps, respectively. Picosecond-level values are typical for free-running solid-state oscillators when integrating down to 1 Hz, because slow thermal and mechanical drifts dominate this range. [10,71] The mixed signal accumulates only 2.6 ps over the same range, corresponding to a reduction by factors of 6.3 and 2.7, respectively. For integration from 10 Hz and from 100 Hz, the reduction is 2.8 to 3.1 for both combs, so the common-mode suppression extends over the full resolvable integration range. The two combs differ by a factor of 2.3 in total timing jitter. This difference accumulates almost entirely below 10 Hz, suggesting that the residual uncorrelated jitter originates from slow, non-common mechanical perturbations outside of the cavity, rather than from fast intracavity dynamics.

## B. Relative Intensity Noise

To study further noise characteristics of the two laser combs, the RIN was measured by reconfiguring the phase noise analyzer to a baseband intensity noise detection. Both the CW and mode-locked (ML) lasing regimes were characterized, with each photodiode output sent directly to the analyzer without filtering or amplification. The results are shown in Fig. 7.

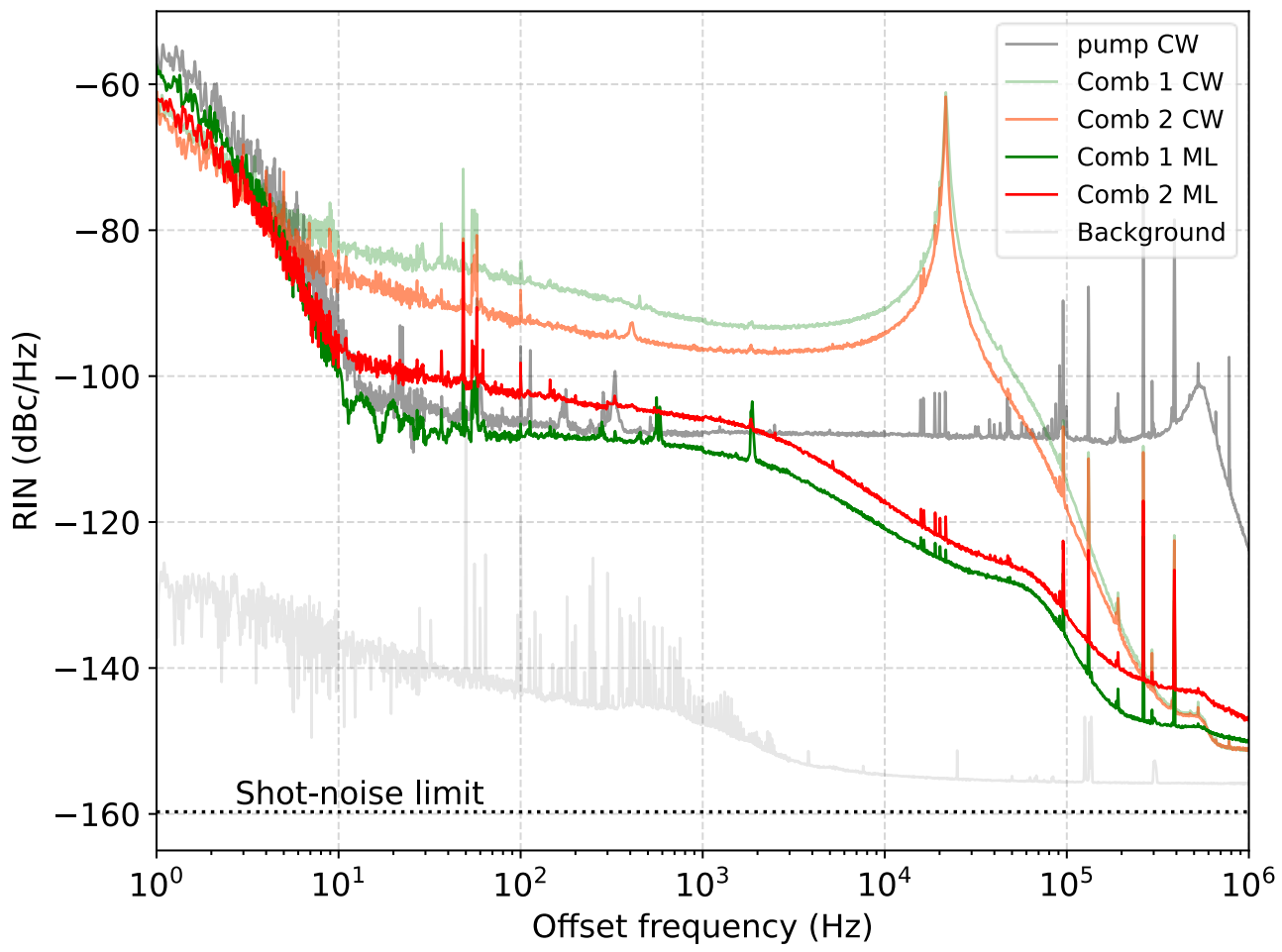

**Fig. 7.** RIN of the two comb outputs in the continuous-wave (light traces) and Kerr-lens mode-locked (dark traces) regimes, measured at baseband without filtering or amplification. The RIN of the pump source, as well as the measurement background, are shown in gray.

In the CW regime, both combs exhibit a pronounced relaxation-oscillation (RO) peak at 20 kHz, reaching −62 dBc/Hz. This peak is completely absent in the ML regime, where the RIN decreases monotonically from −100 dBc/Hz at 10 Hz to below −145 dBc/Hz above 300 kHz. Integrated between 1 Hz to 1 MHz, the RIN falls from 2.83% and 2.03% in the CW regime to 0.08% and 0.10% in the ML regime for the two combs, corresponding to improvement by factors of ~35 and ~20, respectively. We attribute the RO suppression to the fast saturable-absorber action of the soft-aperture Kerr lens mode-locking, which introduces an instantaneous intensity-dependent loss dampening the relaxation-oscillation resonance. A comparable suppression has been reported in Kerr-lens mode-locked Cr:ZnSe lasers [72], where the dominant contribution arises from random quasi-phase-matched second-harmonic generation in the ZnSe polycrystal. Such a mechanism is improbable in our case, as CALGO is centrosymmetric [73] and second-order nonlinear processes are forbidden by symmetry [74]. The RO suppression observed in our Ho:CALGO laser system therefore suggests that the Kerr-lens contribution alone can be sufficient to substantially reduce the RIN of the system. Notably, the relative ordering of the two combs reverses between the two regimes. Comb 1 exhibits the higher integrated RIN in CW operation but the lower value under mode-locking, corresponding to a RO suppression that is 6 dB stronger for Comb 1.

### C. Discussion & Outlook on noise properties of the Ho:CALGO DCS

The noise characterization of the Ho:CALGO DC system showed that the single-cavity geometry yields similar noise parameters and that spurious mechanical noise spikes cancel in the mixed signal. The results of the phase noise measurements showed that the integrated timing jitter drops from 16.3 ps and 7.1 ps to 2.6 ps in a range from 1 Hz to 100 kHz, and the discrete mechanical resonances between 10 Hz and 1 kHz are strongly rejected. For dual-comb spectroscopy, this means the possibility of lowering interferogram-to-interferogram drifts, allowing to use simple methods for phase corrections of interferograms in case of coherent averaging. The RIN characterization results show complete removal of the relaxation-oscillation peak in the ML regime, lowering the integrated RIN from 2.83% and 2.03% to 0.08% and 0.10%.

To date, complete suppression of the relaxation-oscillation peak has not been reported for this class of mode-locked lasers. We note that our previous design of the soft-aperture Kerr-lens mode-locked Ho:CALGO oscillator [48], using the same crystal and pump architecture at a comparable repetition rate, showed a clear peak at ~22 kHz, close to the 20 kHz observed here, which was still present in the ML regime. The main difference between the two systems is the 25% smaller mode radius in the crystal, which raises the intracavity peak intensity by a factor of ~1.8 and strengthens the Kerr lensing. This points to intensity-dependent Kerr-lens loss acting on the laser noise as the likely origin. The long upper-state lifetime of the gain medium additionally reduces the transfer of pump noise to the laser operation. This will be investigated in a follow-up study, focusing on a single comb phase noise and RIN dynamics.

The two combs are not perfectly identical, raising the question about crystal homogeneity and its effect on noise suppression. Since the RIN ordering of the two combs depends on the operating regime, the difference cannot be attributed to the detection arms and must originate in the mode-locking dynamics themselves. During laser operation, we noticed output power, mode-locking threshold, and RIN dependence on the focus position in the crystal volume, which we attribute to inhomogeneous $Ho^{3+}$ doping. The two beams are also laterally separated (2 mm) and therefore experience different local conditions. Local variations of the doping may cause a difference in intracavity peak intensity and thus a stronger effective Kerr nonlinearity for one of the combs, which could explain the reversed RIN ordering. The crystal position used here for laser operation is one of several "sweet spots" where both combs run stably. The residual imbalance measures how far this compromise sits from identical operation.

A gain crystal with more homogeneous doping would directly reduce the asymmetry between the two combs. In addition, a systematic characterization of the gain crystal is required to improve the symmetry between the two combs. Mapping the local absorption, mode-locking threshold and noise as a function of position using would allow the selection of a region in which two laterally separated spots exhibit comparable performance. Reducing the beam separation would increase the shared environment, but it adds risks of gain competition between the two combs. The remaining low-frequency asymmetry in the noise is external and easier to resolve: identical vibration-damped photodiode mounts, and a sealed, passively vibration-isolated laser housing against acoustic, mechanical, and thermal drift.

Overall, the single-cavity geometry provides effective common-mode noise suppression across the 10 Hz to 10 kHz bandwidth. Mode-locked operation yields a low relative intensity noise, free of relaxation oscillation peaks.

## 4. SPECTROSCOPIC BENCHMARKING

To benchmark the DCS system in view of future applications in spectroscopy, we performed a spectral resolution test on single interferograms, without averaging or numerical phase correction, in order to establish the raw performance limits of the free-running dual-comb source. The acquisition parameters for this measurement are set by the two repetition rates. The $f_{rep}$ sets the spectral sampling grid, while $\Delta f_{rep}$ determines the alias-free optical span and the interferogram rate. The value of $\Delta f_{rep}$ was chosen to avoid aliasing while measuring the full spectral range. Alternatively, the spectrum can be band-pass filtered extracavity, which allows higher $\Delta f_{rep}$ and faster acquisition when averaging is applied [75]. The aliasing-free condition $\Delta\nu_{opt} \leq f_{rep}^2/(2\Delta f_{rep})$ limits the optical bandwidth that can be mapped into a single Nyquist zone [1]. In our system, $f_{rep}$ is ~96.5 MHz and our optical bandwidth is 55 nm (3.6 THz) at a central wavelength of 2135 nm. This results in aliasing-limited $\Delta f_{rep}$ values, with an upper limit of 1.3 kHz. We therefore operated with 10x lower order of $\Delta f_{rep}$, operating in the range of ~100 Hz, keeping the down-converted spectrum comfortably inside the aliasing-free window while remaining large enough for good interferogram contrast and straightforward detection with MHz-bandwidth electronics. This yields a resulting down-conversion factor $\alpha = \Delta f_{rep}/f_{rep} \approx 1.0 \times 10^{-6}$, which maps the 3.6 THz optical bandwidth onto an RF band of 3.6 MHz, where it can be recorded with conventional photodiodes and digitizers.

The experimental setup for spectroscopic benchmarking of our dual-comb system is shown in Fig. 8. Two p-polarized laser outputs were sent to the polarizing beam splitter (PBS) through a half-wave plate (HWP) in each path to collinearly overlap the two combs, with minimum power loss. The small residual leakage from both combs was sufficient to track $\Delta f_{rep}$ at the other PBS output.

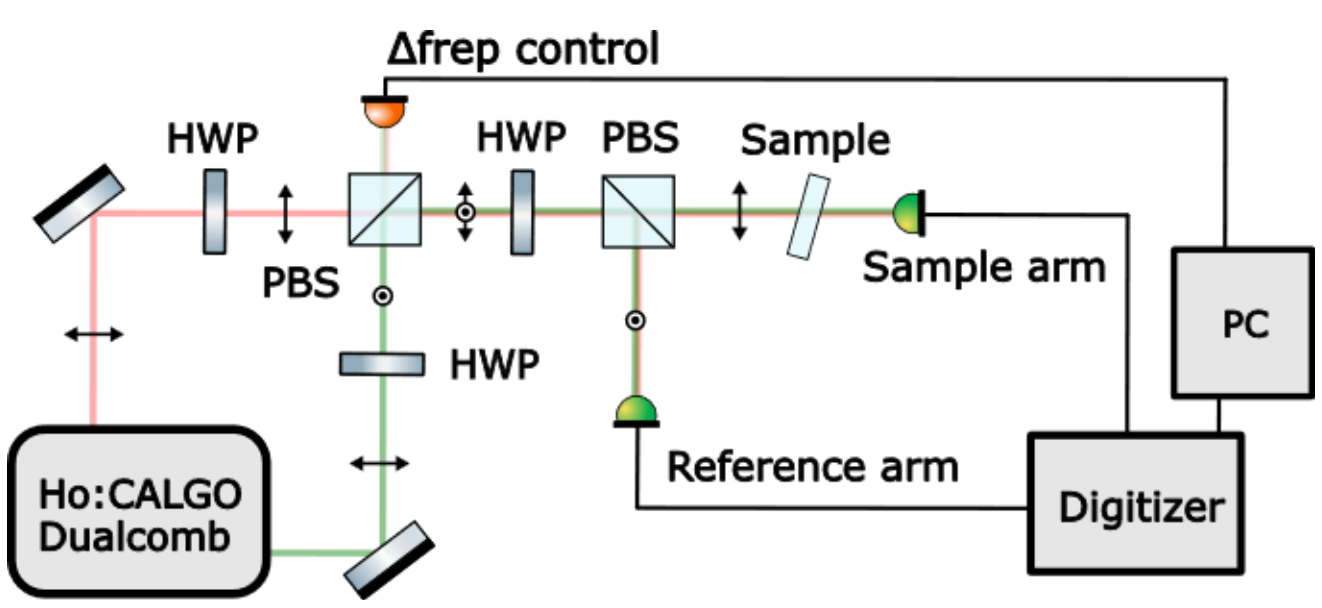

**Fig. 8.** Experimental scheme for DCS benchmarking with plane-parallel plates acting as a Fabry–Pérot interferometer at the sample position.

Next, after passing through another HWP, both combs were sent to the second PBS, splitting the beam path into a sample arm and a reference arm. This scheme minimizes the losses compared to non-polarizing BSs, utilizing the full intensity of both combs for detection in the reference and sample arm. The sample arm included plane-parallel plates, while the reference arm was left as an open path, allowing the modulation imposed by the Fabry–Pérot cavity formed by the plates to be retrieved from the ratio of the two channels. The arms' signals were sent to the extended InGaAs photodiodes and balanced by the last HWP to optimize interferogram (IGM) contrast. Finally, the raw PD signals were captured by a digitizer card and transferred to a PC for post-processing. Figure 9 (a) shows one of the IGM traces recorded without averaging at $\Delta f_{rep}$ = 96 Hz. The recorded time-domain signal was windowed by a Tukey-function and Fourier-transformed to obtain the optical spectrum and compared with the spectrum captured simultaneously using a conventional spectrometer, as shown in Fig. 9(b), showing good agreement.

To test the limits of our free-running single-cavity dual-comb system, we use a physical reference with a known spectral transfer function, in our case using a Fabry–Pérot etalon. This is particularly important in free-running operation if one wants to apply averaging to benefit from high-rate acquisition. Fluctuations of the repetition rate and of the carrier-envelope offset difference between the two pulse trains make the mapping between RF and optical frequencies drift in time [70], so that spectral features progressively broaden and smear. We perform this test to benchmark the performance of our system in a realistic application. This test represents a worst-case scenario, since we applied no averaging or computational phase correction to the interferograms, which would only improve this result.

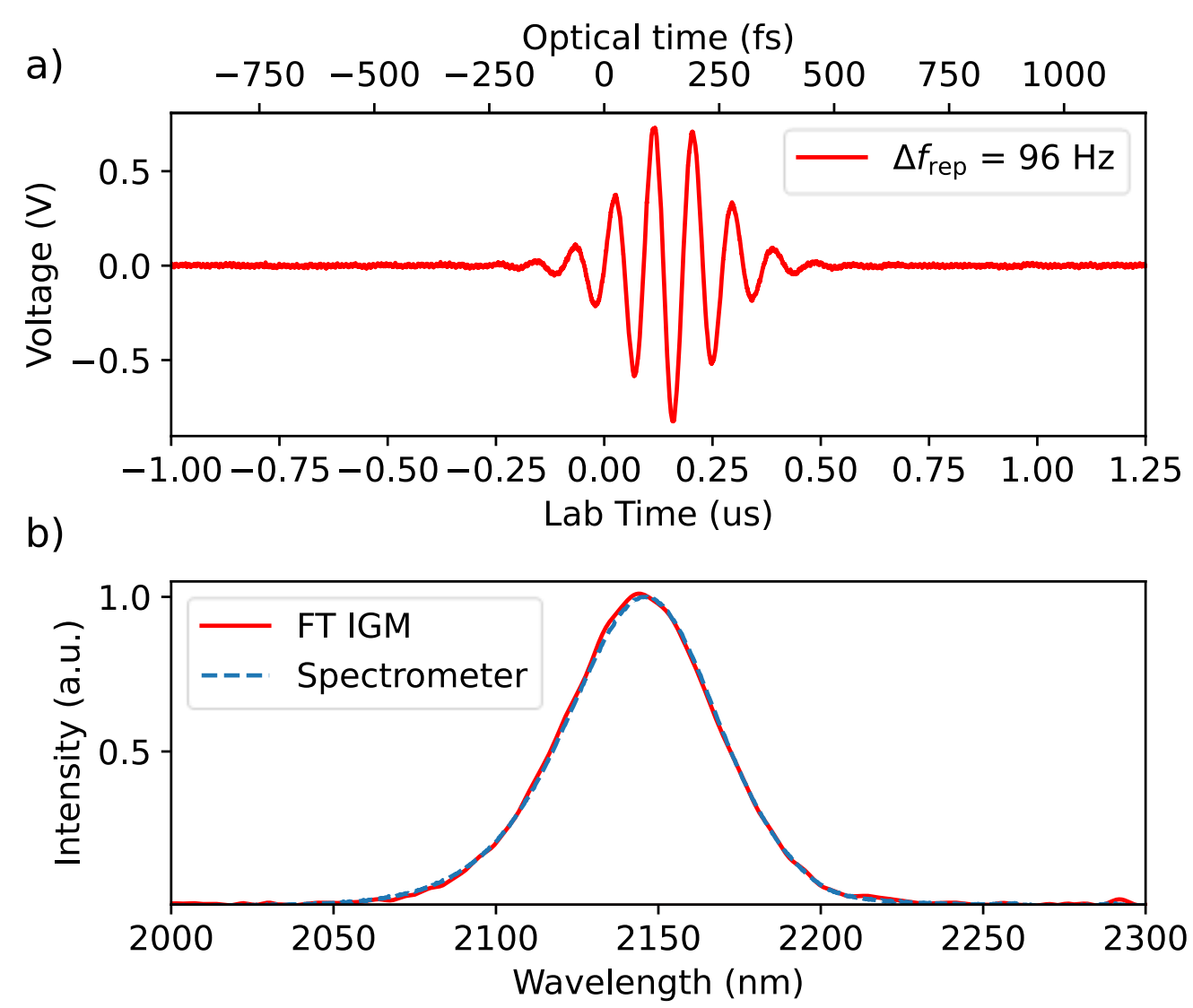

**Fig. 9.** Interferogram (IGM) signal in time domain (a) and resulting spectrum of the laser (b), compared with direct measurement by the spectrometer.

We measured the transmission through 1-, 3-, and 5-mm-thick sapphire plane-parallel plates acting as a Fabry–Pérot interferometer to determine the resolution limitations of our system. These plates induce an absorption feature that can be analytically compared and are chosen to give an idea of how narrow a linewidth our system can resolve without applying additional averaging algorithms.

In Fig. 10(a), the RF interferograms were recorded at $\Delta f_{rep}$ = 134 Hz. This value differs from the one used in the previous measurement because of environmental drifts and the realignment of the setup, but the measurement conditions themselves are identical. The IGM was then converted to the optical domain using the down-conversion factor α, windowed and Fourier transformed, yielding the spectrum shown in Fig. 10(b). The retrieved periodic fringe pattern, shown in Fig. 10(c), is in good agreement with the calculated modulation and with the transmission measured independently by a conventional spectrometer, confirming the accuracy of the optical frequency-axis calibration.

Following calibration, the spectral resolution of the system was evaluated using sapphire plates of increasing thickness. Figure 11 demonstrates the resolution test results. The modulation produced by the 1-mm thick plate, corresponding to a free spectral range (FSR) of 1.08 nm, is clearly resolved. The 3-mm and 5-mm thick plates likewise show a distinct modulation, with FSRs of 0.42 nm and 0.26 nm, respectively, in agreement with the physical and calculated references. In contrast to the 1-mm sample, the traces obtained with the 3-mm and 5-mm thick plates show more pronounced noise spikes, which become more apparent as the modulation period decreases. These spikes presumably have a pure electrical origin and, being uncorrelated with the interferogram signal, are expected to be suppressed by averaging. In the absence of averaging, the single-shot measurements already resolve FSRs as small as 17.1 GHz (0.26 nm). Previous etalon measurements with free-running bulk dual-comb lasers resolved FSRs of 100 GHz and above [10,76,77], in some cases relying on phase correction and coherent averaging. Our single-shot measurement resolves a considerably smaller FSR without any averaging, still remaining far from the limit of 96.6 MHz set by the comb line spacing.

Further improvement of these results requires coherent averaging, which in turn relies on phase alignment of the IGM traces. The residual timing and phase fluctuations between consecutive interferograms can be compensated by time-domain phase correction, e.g., based on a constant-fraction discriminator (CFD) [78]. This approach can enable fast and accurate coherent averaging with improved robustness against repetition-rate and phase fluctuations. Since both combs originate from a common cavity, their mutual coherence is maintained over the entire interferogram duration. Consequently, the correction can be performed entirely in post-processing, without the repetition-rate and carrier-envelope-offset stabilization on a hardware level, required for two independent oscillators. Coherent averaging is expected to suppress the residual noise spikes and to extend the achievable resolution below 10 GHz, as required for detecting narrow-band gas absorption lines.

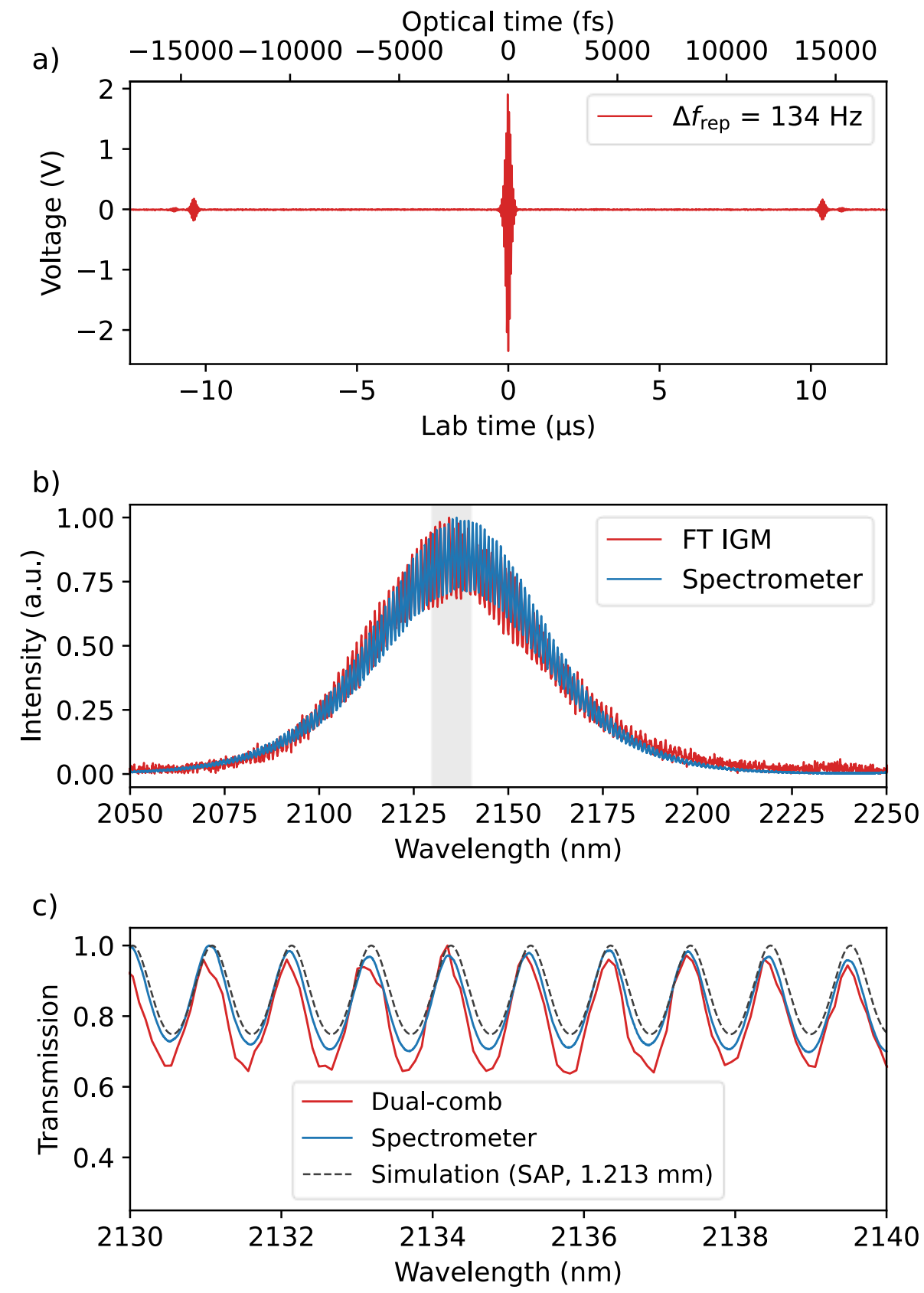

**Fig. 10.** Comparison of Fabry-Pérot etalon effect on the spectra measured with a spectrometer, numerically simulated, and obtained by a dual-comb system. The gray box in (b) demonstrates the zoomed region of (c).

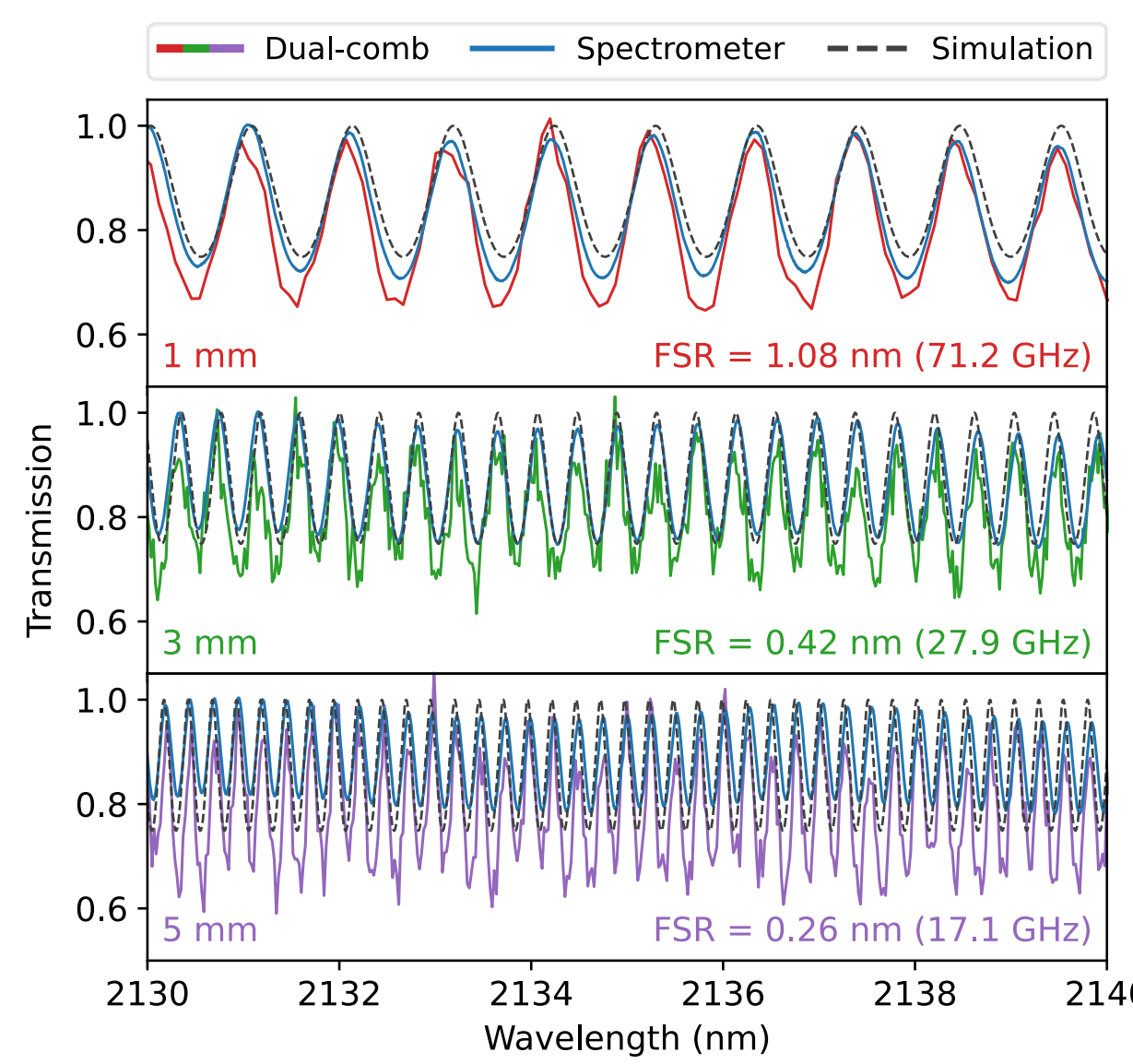

**Fig. 11.** Comparison of 1-, 3-, and 5 mm Fabry-Pérot etalons, acting on the laser spectra. Resulting traces were measured directly with a spectrometer, numerically simulated, and obtained with a dual-comb system. FSR – free spectral range.

## 5. SUMMARY

In summary, we have demonstrated a novel watt-class free-running dual-comb laser operating at a central wavelength of 2.13 µm based on the gain medium Ho:CALGO. The system employs soft-aperture KLM within a spatially multiplexed cavity design, generating two pulse trains with slightly different repetition rates from a single resonator. This approach ensures that both combs share common optical components and experience correlated noise, which is important for maintaining mutual coherence in free-running operation. The laser delivers a total output power exceeding 2 W with sub-100-fs pulse duration and spectral bandwidths exceeding 50 nm, representing an advance in high-power ultrafast single-cavity dual-comb sources at 2.1 µm. Phase noise measurements of the individual combs and of their electrically mixed signal revealed effective common-mode suppression of mechanical noise below 1 kHz, reaching the laser noise floor above 5 kHz. The laser RIN measurements show evidence of relaxation-oscillation suppression in the mode-locked regime, compared with continuous-wave operation.

The benchmarking measurements were performed by collinearly overlapping the two comb outputs on a polarizing beam splitter, and the resulting interference signal was measured with a fast extended-InGaAs photodetector. Wavelength calibration was carried out using 1, 3, and 5 mm plane-parallel sapphire plates acting as a Fabry–Pérot reference, yielding close agreement with the modulation simulations and with complementary spectrometer measurements. The measurements confirm accurate frequency-axis calibration without averaging. The resolution was tested by gradually increasing the thickness of the sapphire plates, reaching a smallest resolvable free spectral range of 0.3 nm. Further decreasing the resolvable FSR is constrained by noise spikes, highlighting the need for a phase-correction algorithm to enable error-free averaging.

Further development will proceed in several directions. Firstly, updates to the cavity mechanical design will be implemented to reduce the fraction of the resonator that is not shared by the two combs, improving the mechanical noise correlations. Secondly, enclosing and purging the cavity with a goal to remove the humidity and temperature dependence that dominates the long-term drift observed in this work. Finally, implementing established interferogram phase-correction algorithms that will allow coherent averaging, improving the raw results reported above without any change to the hardware.

After the aforementioned updates, the high-power 2.1 µm dual comb can be efficiently converted into the THz spectral range using organic crystals [79] , into octave-spanning supercontinua using waveguides [80], and towards MIR using $ZnGeP_2$ (ZGP) crystals [81]. Potential applications include THz-ASOPS techniques for time-domain spectroscopy, mid-infrared molecular fingerprint spectroscopy [82], and rapid greenhouse gas detection for environmental monitoring [17], where no direct driving source in this wavelength has been demonstrated yet.

**Funding.** Deutsche Forschungsgemeinschaft (390677874, EXC2033 RESOLV and 287022738, SFB/TRR MARIE); European Research Council (101138967, Giga2u); Ruhr-Universität Bochum (Open Access Publication Funds),

**Acknowledgment.** We acknowledge support by the DFG Open Access Publication Funds of the Ruhr-Universität Bochum. . The research was carried out in the Research Center Chemical Sciences and Sustainability of the University Alliance Ruhr at Ruhr-University Bochum. Funded by the Deutsche Forschungsgemeinschaft (DFG, German Research Foundation) under Germanys Excellence Strategy – EXC-2033 – Projektnummer 390677874 – RESOLV. The research work presented in this paper was funded by the German Research Foundation ("Deutsche Forschungsgemeinschaft") (DFG) under Project-ID 287022738 SFB/TRR 196 for Project S02. These results are part of a project that has received funding from the European Research Council (ERC) under the European Union's HORIZON-ERC-POC programme (Project 101138967 -Giga2u).

**Disclosures.** The authors declare no conflicts of interest.

**Data availability.** The data underlying the results presented in this paper are available in Ref. 83